\documentclass[prl,twocolumn,preprintnumbers,amsmath,amssymb, nofootinbib]{revtex4-1}
\usepackage{color}
\usepackage[active]{srcltx}
\usepackage{amsmath,amsfonts,amssymb,amsthm,amstext,amscd,eucal,srcltx}
\usepackage{epsfig,graphicx,bm}
\usepackage{epstopdf, epsf}
\usepackage{dcolumn}
\usepackage{hyperref}
\usepackage{cancel}

\newcommand{\mpl}{M_{{}_{\rm {Pl}}}}

\def\AS{{\alpha_k^{{}^S}}}
\def\A{{\alpha_k}}
 \def\B{{\beta_k}}
 \def\AT{{\alpha_k^{{}^T}}}
 \def\BS{{\beta_k^{{}^S}}}
\def\BT{{\beta_k^{{}^T}}}

\def\chis{\chi_{{}_S}}

\newcommand{\be}{\begin{equation}}
\newcommand{\ee}{\end{equation}}
\newcommand{\beqa}{\begin{eqnarray}}
\newcommand{\eeqa}{\end{eqnarray}}
\newcommand{\bi}{\begin{itemize}}
\newcommand{\ei}{\end{itemize}}

\begin{document}
\title{Rescuing Single Field Inflation from the Swampland}
\author{Amjad Ashoorioon$^{1}$}
\bigskip\medskip
\affiliation{$^1$ School of Physics, Institute for Research in Fundamental Sciences (IPM), P .O. Box 19395-5531, Tehran, Iran}
\vfil

\begin{abstract}{The difficulty of building metastable vacua in string theory, has led some to conjecture that in the string theory landscape, potentials satisfy $\left|\nabla V/V\right|\geq c\sim \mathcal{O}(1)$. This condition, which is supported by different explicit constructions, suggests that the EFTs which lead to metastable de-Sitter vacua belong to what is dubbed as {\it swampland}. This condition endangers the paradigm of single field inflation. In this paper, we show how scalar excited initial states cannot rescue single field inflation from the swampland, as they produce large local scalar non-gaussianity, which is in conflict with the Planck upper bound.  Instead, we demonstrate that  one can salvage single field inflation using excited initial states for tensor perturbations, which in this case produce {\it only} large flattened non-gaussianity in the tensor bispectrum. We comment on the possible methods one can prepare such excited initial conditions for the tensor perturbations. }
\end{abstract}


\preprint{\today }
\preprint{IPM/P-2018/072}
\maketitle

\section{Introduction}

With the postulation and formulation of the string theory {\it landscape} \cite{Susskind:2003kw}, the original aspiration for the string theory as a unique quantum theory of gravity, that in the low energy limit reduces more or less uniquely to the Standard Model coupled to gravity, faded. The initial understanding of the landscape was that any local Lorentz-invariant effective field theory, which is semi-classically consistent, can be embedded somewhere in the  infinitely vast array of discrete vacua \cite{Kachru:2003aw} and therefore questions such as the values of cosmological or other fundamental constants reduce merely to some environmental questions for which the detailed string theory construction, at least, at the moment is irrelevant.

Shortly it was realized that not all consistent-looking effective field theories can be UV completed in string theory. For example in \cite{Adams:2006sv}, it was shown that the the sign of the irrelevant operators in such EFTs should be strictly positive, otherwise they will lead to superluminal fluctuations  around non-trivial backgrounds and thus to  IR non-locality and breakdown of causality (please see \cite{Babichev:2007dw} for an opposite perspective on how superluminality can still come across the causality). Also recently, the existence of some issues with the construction of meta-stable de Sitter vacua in String Theory \cite{Maldacena:2000mw, Blaback:2011pn}, has led some to believe that the the string theory landscape is surrounded by a much larger seemingly consistent effective field theories which cannot be UV completed in the string theory, and is dubbed as {\it swampland}. Indeed it is claimed that the string theory landscape is of measure zero in comparison with the swampland \cite{Vafa:2005ui}. The proposition of the swampland conjecture could be regarded as a nostalgia for the glorious days of string theory, from mid-80s to 90s, where the goal of string theory was to uniquely specify the low energy effective field theory.

Several criteria have been presented to distinguish the effective field theories that belong to such a swampland. The first criterion confines the field displacements to be less than $\Delta \sim \mathcal{O}(1)$ in Planckian unites in a theory in which the scalar fields are minimally coupled to gravity \cite{Ooguri:2006in, Agrawal:2018own}. The metric structure on the field space, $G_{ij}(\phi^i)$, can still be non-trivial though. There are evidences from explicit string theory constructions \cite{Lee:2018urn} that as the field range traverses by a large distance $D$, a tower of light modes with mass scale
\be
m\sim \Lambda e^{-\alpha D}\,,
\ee
where $\Lambda$ is the UV cutoff of the theory and $\alpha \sim \mathcal{O}(1)$. Such an exponential lightening of the tower of masses near a particular point could also be understood if it is assumed that the loop corrections drive both gravity and the scalar field to strong coupling at the same energy scale \cite{Heidenreich:2018kpg}. This shows that significant super-Planckian excursions cannot be described by {\it single field} effective field theory. This will have consequences for the {\it large} field inflation if  tensor to scalar ratio $\gtrsim 0.01$ is detected by the CMB B-mode experiments, since in order to realize them, according to the Lyth bound \cite{Lyth:1996im}, field displacements larger than $\mpl$ is required. Such constraint points inevitably in the direction of multi-field inflationary setups to implement the paradigm of inflation in the string theory. Inflationary models like M-flation \cite{Ashoorioon:2009wa, Gauged-Mflation}, Cascade Multi M5 brane inflation \cite{Becker:2005sg, Ashoorioon:2006wc} or N-flation \cite{Dimopoulos:2005ac,Liddle:1998jc} seem to be advantageous from this perspective, although in N-flation all the moduli are lighter than the species UV cutoff \cite{Dvali:2007hz}. In Gauged M-flation \cite{Gauged-Mflation}, however, the hierarchical mass structure of the species allows for only a negligible portion of these modes to remain below the UV cutoff, which is lowered from the Planck mass by the square root of the number of light species \cite{Gauged-Mflation, Dvali:2007hz}. Such many field models of inflation are also privileged to avoid the $\eta$-problem, which is generated from the graviton loop correction \cite{Ashoorioon:2011aa}.   In another approach to realization of multi-field inflation, \cite{Achucarro:2018vey}, the field space metric is nontrivial and curved and thus one instead of following the trajectory dictated by the potential,  follows the geodesic dictated by both the metric of the field space and the potential during inflation. Thus in a finite field space, with a curved trajectory, the dynamical field range can become substantial and super-Planckian.  It seems that even in seemingly single-field models like monodromy inflation \cite{Silverstein:2008sg, Marchesano:2014mla}, which were proposed in the context of string theory to achieve single field super-Planckian field displacements, without considering such multi-field effects,  one would violate this swampland criterion \cite{Landete:2018kqf}. The claim of \cite{Achucarro:2018vey} however is that with such curved trajectory, one would reduce the speed of sound for the fluctuations which is determined by the ratio of rate of turning of the inflationary trajectory and the mass of the heavy field.

In absence of any observation of B-modes, there is no need to have super-Planckian excursions in the field space to realize inflation. The small field inflationary models could still serve the purpose quite well. Models like $\mathbb{K}{\rm L}\mathbb{M}{\rm T}$ will be a model in the context of string theory that can work \cite{Kachru:2003sx}. However the setup of the model is based on the uplift of AdS vacuum to the de-Sitter vacuum which was doubted to work in several papers \cite{Blaback:2011pn} (In the literature, there are several other approaches to construct de Sitter vacuum from AdS using nilpotent superfield \cite{Kallosh:2018wme}). Indeed such studies led the authors of  \cite{Agrawal:2018own} to propose the second swampland criterion which not only proscribe the existence of a (metastable) de-Sitter minimum but also puts a lower bound on how close to flat the slope of a potential can be in the string theory landscape,
\be
\frac{\left|\nabla V\right|}{V}\geq c\,,
\ee
where $c$ is suggested to be $\mathcal{O}(1)$.  The lowest value for $c$ is obtained for a construction, which involves ${\rm O4}$ planes and no D-brane source, with the value of $c_{{}_{\ast}}=\sqrt{2/3}\sim 0.8$. Such construction although would allow for accelerated expansion, something that needs $c<\sqrt{2}$ in 4D, will lead to large tensor to scalar ratio, $r$, in a single field inflationary model, which is expected to be $r=8 c^2$ from the single field consistency relation. With the current bound on  $r$, reported by Planck 2018 data \cite{Akrami:2018odb}, and having a single field inflation, one would need $c<0.0894$ which is not compatible with the smallest value of $c$ reported in \cite{Obied:2018sgi}, $c_{{}_{\ast}}=0.8$. Here we take the relation $c\sim \mathcal{O}(1)$ at the face value and assume that $c\geq c_{{}_{\ast}}$ as the best case scenario.  In order to surpass the limit the maximum experimental bound set on $r$, one alternatively can use the K-inflation mechanism \cite{ArmendarizPicon:1999rj} in which the consistency relation is modified as $r=8 c_{{}_S} c^2$, where $c_{{}_S} $ is the scalar sound speed. With $c_{{}_S} \leq 1$, one can reconcile the small upper bound set on $r$ and relatively large minimum of $c_{{}_{\ast}}$ reported in \cite{Obied:2018sgi} . An example of such an attempt in string theory is DBI inflation \cite{Alishahiha:2004eh}.  However for DBI and (K-inflation in general), the level of non-gaussianity, which is mainly in the equilateral shape, is \cite{Chen:2006nt}
\be
f_{{}_{\rm NL}} ^{{}^{\rm equil}}\simeq \frac{35}{108} \left(-1 + \frac{1}{c_{{}_S}^2}\right)\,.
\ee
The latest report of the Planck on the equilateral shape of non-gaussianity is \cite{Ade:2015ava}
\be
f_{{}_{\rm NL}}^{{}^{\rm equil}}=-4\pm 40~~~(68\%{\rm C.L.})\,.
\ee
To remain in the $2\sigma$ limit of Planck 2015 results, one needs \footnote{Please see \cite{Kinney:2018nny} for a similar bound.}
\be
c \lesssim 0.35\,,
\ee
which is again smaller than $c_{{}_{\ast}}=0.8$ \footnote{In \cite{Achucarro:2018vey}, where due to the curved field space the speed of sound is reduced form one, there seems to be a similar conflict with large non-gaussianity from observation for rather large values of $c$.}.  

Another suggestion for having a CMB-compatible single field inflation with steep potential has been proposed using the curvaton mechanism \cite{Kehagias:2018uem}. Since the framework suggests an ingredient beside the single field inflaton, {\it i.e.} the curvaton, to imprint the perturbations on the CMB, we consider such a mechanism to be a multi-field setup. In the curvaton scenario, substantial amount of local non-gaussianity is generated, that needs to be checked against the Planck latest upper limit, noting that the swampland criteria should be applied to the curvaton potential too. Alternatively, one can use the idea of warm inflation to reconcile the paradigm of inflation with the swampland conjecture \cite{Motaharfar:2018zyb}, although again due to the necessity of the existence of a thermal bath of other species that the inflaton has to interact with in order to slow down on a steep potential, we categorize the model as a multifield inflationary setup.  Inflation on the brane \cite{Lin:2018kjm}, in which there are corrections to the right hand side of the Friedmann equation proportional to the density squared, has also been suggested to realize inflation despite the swampland criteria \footnote{In \cite{Matsui:2018bsy} and \cite{Dimopoulos:2018upl}, it was shown that the swampland conjecture is not compatible with the idea of eternal inflation. The refined version of the swampland conjecture \cite{Ooguri:2018wrx}, nonetheless, seems to be able to accommodate eternal inflation \cite{Kinney:2018kew}.}.

As it has been emphasized frequently in the past, the  predictions of the inflationary models for the CMB temperature anisotropy
depend on the  initial state of the quantum fluctuations as well as the details of the inflationary potential. Although it  is usually assumed that the  fluctuations start off in the Bunch-Davis (BD) vacuum \cite{Bunch:1978yq}, there could be various mechanisms to excite them from the BD vacuum when they leave the horizon \cite{Initial-data-literature}. Sometimes the excited initial condition impose themselves as the only initial conditions for which a theory would yield finite observables. This is observed recently in the context of the general Extended Effective Field Theory of Inflation \cite{Ashoorioon:2018ocr}, where the theory exhibits breakdown of unitarity at very short wavelengths. The effect of such initial states on the observables have been extensively studied in the literature \cite{Chen:2006nt,Ashoorioon:2010xg}. In \cite{Ashoorioon:2013eia}, it was noticed that, for the region of parameter space in which the scalar perturbations start from highly excited initial conditions  --  what we dubbed as super-excited initial states -- no matter what the initial state of tensor perturbations is, $r$ is suppressed.  Recently in the context of swampland conjecture, such excited initial conditions were offered \cite{Brahma:2018hrd, Das:2018hqy} as a resolution to the large $r$  predicted by the inflationary potentials with large slopes, which are suggested to be the only ones that could be realized within the string theory landscape. However as we will explain, for such potentials, the predicted local non-gaussianity will be large and violates the latest Planck upper bounds, if the scalar initial conditions are excited \cite{Ade:2015ava}.

The outline of the paper is as follows: first, we  show how with excited initial conditions for perturbations, one can modify $r$ predicted by the model. We explicitly demonstrate that scalar excited initial conditions are unable to assist us toward this goal, if $c\gtrsim 0.21$. We demonstrate how tensor super-excited states are capable in both reducing or increasing $r$, depending on the phase difference between the tensor Bogolyubov coefficients. We spell out different signatures of such tensor super-excited initial conditions  in the tensor bispectrum. We conclude the paper finally and point out directions for further probes of the proposal.

\section{Suppressing the Tensor to Scalar Ratio with Excited States}

We have briefly reviewed the formalism of cosmological perturbation theory with excited initial condition in the Appendix.  For general excited initial conditions for the  tensor and scalar perturbations, the power spectra take the form,
\beqa\label{power-spectra}
{\mathcal P}_{{}_S}={\mathcal P}_{{}_{\rm BD}}^{{}^S}\,\gamma_{{}_S} \,,\\
 {\mathcal P}_{{}_T}={\mathcal P}_{{}_{\rm BD}}^{{}^T}\ \gamma_{{}_T}\,,
\eeqa
where
\beqa
{\mathcal P}_{{}_{\rm BD}}^{\rm S}&=&\frac{1}{8\pi^2\epsilon}\left(\frac{H}{\mpl}\right)^2,\qquad \gamma_{{}_S}=|\AS-\BS|^2_{{}_{k={\cal H}}}\,,\\
{\mathcal P}_{{}_{\rm BD}}^T&=&\frac{2}{\pi^2}\left(\frac{H}{\mpl}\right)^2\,,\qquad~~ \gamma_{{}_T}=|\AT-\BT|^2_{{}_{k={\cal H}}}\,,
\eeqa
The tensor to scalar ratio, $r$, is then given by
\begin{equation}\label{consistency}
r\equiv\frac{{\mathcal P}_{{}_T}}{{\mathcal P}_{{}_S}}=16 \gamma\epsilon,\quad \gamma=\frac{\gamma_{{}_T}}{\gamma_{{}_S}}=\left|\frac{\AT-\BT}{\AS-\BS}\right|^2\,.
\end{equation}
Following \cite{Ashoorioon:2013eia}, let us parameterize the Bogolyubov coefficients $\alpha,\ \beta$ for tensor and scalar perturbations as follows,
\be\label{parametrization}
\begin{split}
\alpha^{{}^S}_k =\cosh\chi_{{}_S} e^{i\varphi_{{}_S}}\,&,\quad \beta^{{}^S}_k =\sinh\chi_{{}_S} e^{-i\varphi_{{}_S}}\,,\\
\alpha^{{}^T}_k =\cosh\chi_{{}_T} e^{i\varphi_{{}_T}}\,&,\quad \beta^{{}^T}_k =\sinh\chi_{{}_T} e^{-i\varphi_{{}_T}}\,.
\end{split}
\ee
With this parametrization, $\chi_{{}_{S}}\simeq\sinh^{-1}\beta_0^{{}^{S}},\ \chi_{{}_{T}}\simeq\sinh^{-1}\beta_0^{{}^{T}}, \ e^{-2\chi_{{}_S}}\leq \gamma_{{}_S}\leq e^{2\chi_{{}_S}}$,  and $e^{-2\chi_{{}_T}}\leq \gamma_{{}_T}\leq e^{2\chi_{{}_T}}$, where $\gamma_{{}_S}$ and $\gamma_{{}_T}$ are the modulations of the power spectrum with respect to the power spectra obtained with Bunch-Davies vacuum initial conditions. $\beta_0^{{}^S}$ and $\beta_0^{{}^T}$, as we will see, are the norms of the second Boglyubov coefficients of, respectively, the scalar and tensor perturbations, to which  the inflationary modes are pumped to when their physical wavelengths pass through the new physics hypersurfaces. From $\nabla V/V=c$, one obtains $\epsilon= \frac{c^2}{2}$. Also from the expression for the scalar spectral index, $n_{{}_S}=1+2\eta-6\epsilon$, and the mean of its measured value $ n_{{}_S}= 0.9649$ from the Planck 2018 results \cite{Akrami:2018odb}, one can find an estimate for the second slow-roll parameter, $\eta\simeq -0.01755-\frac{3 c^2}{2}$. With $c\sim \mathcal{O}(1)$, such an $\eta$ parameter is large (please see \cite{Garg:2018reu} too), but inflating with such a large $\eta$ parameter is not impossible \cite{Kinney:2005vj}. The only requirement for having an inflationary background is having $\epsilon<1$, which is  what we have assumed so far by setting  $c<\sqrt{2}$.

We also consider a model for the excitation of the scalar and tensor modes suggested in \cite{Boyanovsky:2006qi} \footnote{It could also be any function for which $|\beta_k|^2$ falls off as $k^{-(4+\delta)}$ with $\delta >0$ to ensure the Hadamard condition \cite{Holman:2007na}. The details of such a functional behavior does not affect the conclusions of this paper.},
\beqa\label{betas-Gaussian}
\beta^{{}^S}_k & \propto & \beta^{{}^S}_{{}_0} \exp\left\{{{-k^2/\left[ a(\tau)M_{{}_S} \right]}^{2}}\right\}\,,\\
\beta^{{}^T}_k & \propto & \beta^{{}^T}_{{}_0} \exp\left\{{{-k^2/\left[ a(\tau)M_{{}_T} \right]}^{2}}\right\}\,,
\eeqa
 which realizes the picture in which the modes get excited almost uniformly as their physical momenta pass through the new physics hypersurface. This picture of excitation could be realized, for example, with modified dispersion relations, where the last inflationary mode that exit the horizon at the end of inflation goes through all different phases of modified dispersion relation \cite{Ashoorioon:2011eg,Ashoorioon:2017toq}. Large occupation numbers for the states could be realized with the ones with a period of negative slope in an interim sub-horizon phase at high momenta in the dispersion relations \cite{Ashoorioon:2017toq}. Such a picture could be realized in the Extended EFT of inflation \cite{Ashoorioon:2018uey}, although in that framework, the only modification imparts upon the tensor perturbations  is a change in the speed of propagation. In order to modify the dispersion relation of tensor perturbations, one in principle should consider alternative pictures in which the gravity is modified in infinite UV. An example of such a behavior could be manifested  in  Ho\u{r}ava gravity \cite{Horava:2009uw, Bogdanos:2009uj} where the dynamical exponent evolves from $z=3$ ($\omega^2\propto k^6$) in the UV to ($\omega^2\propto k^2$) $z=1$ in the IR.  The above form roughly implies
that the non-BD state becomes important as the physical momentum of the scalar and tensor modes, respectively, redshift and pass through  $M_{{}_S}$ or $M_{{}_T}$. The new physics, which is presumed to have a description in terms of a generally covariant  effective field theory, will affect scalar and tensor sectors differently.
That is the reason we have assumed that the scale of new physics for scalar and tensor perturbations are in general different, although they can be identical too. The $\alpha_{{}_0}, \beta_{{}_0}$ parameters are also not necessarily the same for the scalar and tensor modes.

The factor $\gamma$  which parameterizes the effects of initial states can in principle be bigger or smaller than one. If $\gamma$ is smaller than one, it will help suppress
$r$. This is what required to reconcile \cite{Ashoorioon:2013eia} single field inflationary models like $m^2\phi^2$ that predicts large $r$ with the Planck data \cite{Planck:2013jfk}.

Now from the normalization of the power spectrum at the pivot scale, $k=0.002~{\rm Mpc}^{-1}$, $\mathcal{P}_{{}_S}\simeq 2.0989\times 10^{-9}$, one obtains
\be\label{H-gammas}
\frac{H}{\mpl}\simeq \frac{1.14838\times 10^{-4} c}{\sqrt{\gamma_{{}_S}}}\,,
\ee
and the backreaction constraint for the scalar perturbations, eq. \eqref{beta-scalar-tensor-backreaction},
constrains $M_{{}_S}/H$ as
\be
\frac{M_{{}_S}^2}{H^2}\lesssim \frac{4353.955 \sqrt{\gamma_{{}_S}} }{\sinh \chi_{{}_S}} \sqrt{3c^2+0.0351}\,,
\ee
As noticed in \cite{Ashoorioon:2013eia}, for large occupation number in the scalar sector, $\chi_{{}_S}\gg 1$, $\sqrt{\gamma_{{}_S}}\sim \sinh\chis$,~$\varphi_{{}_S}\simeq \pi/2$ and
one obtains an upper bound on how large $M_{{}_S}$ can be with respect to $H$. In such a sector of parameter space, irrespective of the
modifications in the tensor spectral index the ratio $\gamma\lesssim 1$ \cite{Ashoorioon:2013eia}.
In principle with this, one could satisfy the constraints $r<0.065$ from the Planck 2018 data, even if $c\sim \mathcal{O}(1)$. For example for no modification in the initial condition for the tensor perturbations,
$\gamma_{{}_T}\simeq 1$, one can suppress
the tensor to scalar ratio, $r=8c^2$, by a factor of $\exp(-2\chi_{{}_S})$. For sufficiently large values of $\chi_{{}_S}\gg 1$, one could reduce $r$ and bring it down to a value which is compatible with the Planck 2018 upper bound even if $\mathcal{O}(1)$. For example for $c=0.8$, $\chi_{{}_S}\sim 2.18$ to be compatible with the latest Planck upper bound on the amplitude of tensor perturbations.  Larger values of $c$, require  more populated excited states.

As explained before, however, this part of parameter space suffers from too large value for the local non-gaussianity in the scalar sector. The upper bound on local non-gaussianity from Planck 2015 data is $f_{{}_{\rm NL,S }}^{{}^{\rm local}}\leq 10.8$ with $95\%$ C.L \cite{Ade:2015ava} . The local non-gaussianity in a super-excited state is given by \cite{Ashoorioon:2013eia, Ashoorioon:2015pia, Ashoorioon:2016lrg}
\be
f_{{}_{\rm NL, S}}^{{}^{\rm local}}=\frac{5}{3}\frac{c^2}{2} \frac{k_{{}_S}}{k_{{}_L}}\,,
\ee
where $k_{{}_S}$ is the shortest wavelength probed by Planck, $\ell\sim 3000$, and $k_{{}_L}$ is the longest wavelength at which the cosmic variance is limited, $\ell\sim 10$. With the current bound on the local non-gaussinity  from the 2015 data, the bound on $c$ is
\be
c\lesssim 0.21\,,
\ee
which is still small compared to the smallest slope realized from various stringy setups \cite{Obied:2018sgi}.

There is an overlooked part of parameter space in \cite{Ashoorioon:2013eia} which can help us solve the above tension from the non-gaussianity bound for the scalar perturbations. In order to avoid the above conflict with the local non-gaussianity bound for the scalar perturbations let us assume that $\beta_{{}_0}^{{}^{S}}\simeq 0$,  for scalar perturbations, {\it i.e.} the scalar modes start from the Bunch-Davies vacuum and therefore $\gamma_{{}_S}=1$. This would remove the tension from the non-gaussianity bound. It also keeps the Hubble parameter the same as what is obtained by the Bunch-Davies vacuum. Now let us try to see if one can get the suppression in $r$ by starting off from the excited state in tensor perturbations. Since in our analysis the tensor initial state parameters $\alpha^{{}^T}, \beta^{{}^T}$ and those of
the scalars are taken to be independent, we should in principle make sure that the
the smallness of the backreaction, i.e. \eqref{delta-rho-free} and \eqref{background-backreaction}, for the tensor modes. The backreaction constraint for the tensor perturbations dictates
\be
\beta_{{}_0}^{{}^T}=\sinh \chi_{{}_T}\leq \sqrt{\epsilon\eta}\frac{H M_{{}_{\rm Pl}}}{M_{{}_T}^2}\,.
\ee
Using \eqref{H-gammas} to eliminate $\mpl$ from the above equation one obtains
\be\label{MT-H}
\frac{M_{{}_T}^2}{H^2}\lesssim \frac{4353.95\sqrt{3c^2+0.0351}}{\sinh\chi_{{}_T}}\,.
\ee
Contrary to the case for the scalar perturbations, one cannot increase the number of particles in the super-excited tensor perturbations, since then the EFT for tensor perturbations breaks down assumption, {\it i.e.} $M_{{}_T}$ becomes smaller than $H$. In fact in order to avoid that, there should be an upper bound on the number of excited states in the tensor perturbations state. With $c=c_{{}_{\ast}}=0.8$,
\be
\chi_{{}_T}^{{}^{\cancel{\rm EFT}}}\sim 9.41\,.
\ee
In this part of the parameter space, the factor $\gamma$ which modulates the RHS of the consistency condition is given by
\be
\gamma=\gamma_{{}_T}=e^{2\chi_{{}_T}}\sin^2 \varphi_{{}_T}+e^{-2\chi_{{}_T}}\cos^2 \varphi_{{}_T}\,.
\ee
One notes easily that in this part of the parameter space, the tensor to scalar ratio can either be suppressed or enhanced depending on the phase difference between the Bogoliubov coefficients and the number of particles in the super-excited state. If $\tan^2 \varphi_{{}_T}<e^{-4\chi_T}$, the tensor to scalar ratio will decrease with respect to the vanilla result, {\it i.e.} $\gamma< 1$, otherwise it will enhance it, $\gamma>1$. In order to suppress $r$, from the value the single field consistency suggests, $8 c^2$, we stick to the first option. If $\varphi_{{}_T}\sim 0$, then with $c=c_{{}_{\ast}}=0.8$, one obtains $\chi_{{}_T} \sim 2.183$ which is smaller than $\chi_{{}_T}^{{}^{\cancel{\rm EFT}}}$ for which the EFT breaks down. In fact for such a value of $\chi_{{}_T}$, the scale of new physics for tensor perturbations can go up to
\be\label{MT-H-1.067}
M_{{}_T}^{{}^{\rm max}}\simeq 37.28 H \,,
\ee
which provides enough  hierarchy between the scale of new physics and the inflationary Hubble scale, $H$.

\section{Tensor Bispectrum with Excited Tensor Initial Conditions}

Having excited initial condition for tensor perturbations, we expect to see enhanced non-gaussian signatures in the tensor bispectrum. In particular, like scalar fluctuations with excited initial conditions, one can show that the non-gaussianity will be enhanced for the flattened, $k_1+k_3\approx k_2$ \cite{Chen:2006nt} and local, $k_3\ll k_1\approx k_2$ \cite{Agullo:2010ws}, configurations. It can be shown that the enhancement for the flattened configuration is \cite{Ashoorioon:2018soon}
\beqa
&&h_{{}_{\rm NL}}^{{}^{\rm flat}}\simeq \left[\frac{k_1 k_2 k_3 (k_1^2+k_1 k_3+k_3^2)}{8\left[k_1^3+k_3^3+(k_1+k_3)^3\right]}  \left(-\frac{1}{2}e^{4\chi_{{}_T}} \sin^2\varphi_{{}_T} \times \right.\right. \nonumber\\
&& \left. \left. (2+\cos2\varphi_{{}_T})+\frac{1}{4}(3-\cos 4\varphi_{{}_T})+\frac{1}{4} \cos^2 \varphi_{{}_T}(-2+\cos 2\varphi_{{}_T}) \right. \right. \nonumber \\
&&\left. \left. \times e^{-4\chi_{{}_T}} \right) \tau_i^2-\frac{k_1^2+k_1 k_3+k_3^2 }{\left[k_1^3 + k_3^3 + (k_1 + k_3)^3\right]}\times \tau_i  \right.\nonumber\\
&& \left. \left( e^{4 \chi_{{}_T}} \cos\varphi_{{}_T} \sin^3\varphi_{{}_T}+\frac{1}{4}\sin 4\varphi_T-e^{-4 \chi_{{}_T}} \cos^3\varphi_{{}_T}  \sin \varphi_{{}_T}\right)\right.\nonumber\\
&& \left. +\mathcal{O}(\tau_i^0)\frac{}{}\right]\times(e^{2\chi_{{}_T}}\sin^2\varphi_{{}_T}+e^{-2\chi_T{{}_T}}\cos^2\varphi_{{}_T})^{-2}\nonumber\\
\eeqa
where $\tau_i$ corresponds to the initial conformal time where the smallest of $k_i$'s, which in this case we assume to be $k_1$, passes through the new physics hypersurface,
\be
\tau_i\equiv-\frac{M_{{}_T}}{k_1 H}\,.
\ee
An interesting point about the above result is that, the non-gaussianity in the flattened configurations for an excited initial state for tensor perturbations is proportional to $\tau_i^2$ in the leading order, and not $\tau_i$ as in  the case of excited scalar perturbation in slow-roll inflation. This is in contrast with the excited initial state for scalar perturbation for the slow-roll inflation in which the enhancement was proportional to $\tau_i$ \cite{Chen:2006nt}. Even if one power of  $\tau_i$ is lost after the projection on the 2-dimensional CMB surface \cite{Holman:2007na}, the amplitude of non-gaussianity for tensor perturbation in the flattened configuration could still be large and detectable. For large values of $\chi_{{}_T}$ and for $0\ll\varphi_{{}_T}\ll\pi$, similar to the case of scalar perturbations with excited initial mode, the dependence on the excitation amplitude of the excited mode would drop out. However as we noted above, in order to make $\gamma\ll 1$, one needs to focus on the limit in which  $\tan^2 \varphi_{{}_T}<e^{-4\chi_{{}_T}}$, {\it i.e.} $\varphi_{{}_T}$ is very close to zero. In this case the $\chi_{{}_T}-$dependent factor in the amplitude of non-gaussianity is enhanced if $\chi_{{}_T}$ is increased. In the case of tensor perturbations, there is of course, a limit on how much we can increase $\chi_T$ before reaching the onset of break down of EFT, $\chi_{{}_T}<\chi_{{}_T}^{{}^{\cancel{\rm EFT}}}$. Also with increasing $\chi_{{}_T}$, the $M_{{}_T}/H$ factor can not be enhanced arbitrarily. For $c=c_{{}_{\ast}}=0.8$ and $\chi_{{}_T} \sim 2.18$, which was required to make the inflationary potential coming from the swampland compatible with Planck 2018 data, the maximum momenta-dependent amplitude for the flattened configurations is
\be\label{hnl-mom-dep}
h_{{}_{\rm NL}}^{{}^{\rm flat, max}}\simeq 1638.48 \frac{k_2 k_3 (k_1^2+k_1 k_3+k_3^2)}{8 k_1\left[k_1^3+k_3^3+(k_1+k_3)^3\right]}\,,
\ee
where we have assumed that one factor of $\tau_i$ will be lost after projection on the 2 dimensional CMB surface \cite{Holman:2007na, Holman:2007na}.  Above we have assumed that $M_{{}_T}/H$ takes its maximum value given by eq. \eqref{MT-H} or for the case of $\chi_{{}_T} \sim 2.18$, $M_{{}_T} /H\sim 37.28$. For $k_1=k_3=k_2/2$,
\be
h_{{}_{\rm NL}}^{{}^{\rm flat, max}}\simeq  122.89\,.
\ee
We should emphasize that this is the maximum non-gaussianity expected in the flattened configurations, when $\chi_{{}_T} \sim 2.18$. In principle $M_{{}_T}/H$ could be less than its maximum value, \eqref{MT-H-1.067}, and then the predicted amplitude of the non-gaussianity in the flattened configuration will decrease. Also we should stress the slight momenta dependent nature of the non-gaussianity in the flattened configuration. Above we have assumed that the smaller momenta, $k_1$ and $k_3$ are about half of the largest one, $k_2$. By reducing the magnitude of one of the momenta, the value of this momenta-dependent factor increases. For example, for $5 k_1=\frac{5}{4} k_3=k_2$,
\be
h_{{}_{\rm NL}}^{{}^{\rm flat, max}}\simeq 452.74  \,.
\ee
Making $k_1$ smaller in the flattened configuration would not probably make sense in the flattened configuration, since then we would approach the local one which is discussed and analyzed below.

Besides the flattened configurations, it can be shown that the local configurations will be enhanced too. In the configuration $k_3\ll k_1\approx k_2$, the
\beqa\label{hnl-mom-dep}
&&h_{{}_{\rm NL}}^{{}^{\rm local}}\simeq -\frac{k_1}{4k_3}  \left(e^{4 \chi_{{}_T}} \cos\varphi_{{}_T} \sin^3\varphi_{{}_T}+\frac{1}{4}\sin 4\varphi_{{}_T}\right. \nonumber\\
&&\left. -e^{-4\chi_{{}_T}}\cos^3\varphi_{{}_T}\sin \varphi_{{}_T}\right) (e^{2\chi_{{}_T}}\sin^2\varphi_{{}_T}+e^{-2\chi_{{}_T}}\cos^2\varphi_{{}_T})^{-2}\nonumber\\
\eeqa
Although there is an enhancement proportional to the ratio $k_1/k_3$ in general for the excited tensor modes, for the ones  that suppress $\gamma$,  and reconcile the single field inflation with the swampland idea, the enhancement is lost as $\tan\varphi_{{}_T}\ll e^{-4\chi_{{}_T}}$. Therefore we conclude that one should not see an enhanced local configurations in the three-point functions for tensor perturbations.

\section{conclusion}

In this work, we tried to realize single field inflationary models compatible with the latest Planck data, taking into account the bound swampland criterion  sets on the steepness of the potentials that can be realized within the stringy setups. We showed that excited initial conditions for scalar perturbation will enhance the amplitude of scalar local non-gaussianity and drive it beyond the upper limit set by the Planck experiment. We came up against this by super-excited tensor initial conditions which can suppress $r$ too. In fact, with such excited initial conditions for tensor perturbations, $r$ can be both lowered or enhanced, depending on the phase difference between the first and second tensor Bogolyubov coefficients. The enhancement, which can be obtained for rather large $\tan \varphi_{{}_T} \gg e^{-4\chi_{{}_T}}$,  could be used to obtain large $r$'s from sub-Planckian field excursions. In this work, however, our goal was to reduce $r$, which we achieved by having a highly populated tensor initial conditions and small values of $\varphi_{{}_T}$. The scalar perturbations were assumed to begin from B.D. vacuum. In this region of the parameter space, of course, there will be a limit on how large the number of particles in the tensor initial states could be otherwise one will violate the EFT by pushing the scale of new physics for tensor perturbation evolution, $M_{{}_T}$, above the inflationary Hubble parameter. Similar to the case of scalar perturbations with excited initial conditions, one would expect enhancement for the flattened and local configurations in the tensor bispectrum. For the peculiar case of tensor excited initial conditions which were required to suppress $r$, the local enhancement will be indeed quite negligible. However for the flattened configurations can reach $h_{{}_{\rm NL}}^{{}^{\rm flat}}\sim {\rm few}\times 100$. In this work we only focused on the tensor bispectrum. However in principle with such super-excited tensor modes, the scalar-scalar-tensor bispectrum is expected to cget modified too. The signature of the model for such correlations is something that we postpone to the upcoming article \cite{Ashoorioon:2018soon}.


\section{Acknowledgment}

I am thankful to O. Lechtenfeld and the Riemann center for their hospitality when this work started. I am also grateful to  G. Shiu  for insightful comments on the draft.

 \appendix
\section{Appendix 1}\label{Apndx1}
Let us consider a scalar field minimally coupled to gravity and look at the scalar and tensor perturbations in this setup with general excited initial conditions. The equation of motion for the scalar and tensor perturbations, respectively, are
\beqa\label{u-p-eq}
u^{\prime\prime}_k+\left(k^2-\frac{z^{\prime\prime}}{z}\right)u_k=0\,, \\
p^{\prime\prime}_k+\left(k^2-\frac{a^{\prime\prime}}{a}\right)p_k=0\,,
\eeqa
where
\beqa\label{u-mukhanov}
u=-z \left( \frac{a^{\prime}}{a}\frac{\delta \phi}{\phi^{\prime}}+\Psi\right), && \quad z\equiv \frac{a \phi^{\prime}}{\cal H}, \quad {\cal H}\equiv \frac{a^{\prime}}{a}\,,\\
&&p_{ij}=a h_{ij}\,,
\eeqa
$u$ is the gauge-invariant Mukhanov-Sasaki variable \cite{Mukhanov:1981xt, Kodama:1985bj} which is a combination of the scalar perturbations of the metric $\Psi$ and the fluctuations of the inflaton $\delta \phi$ and $a$ is the scale factor. The prime denotes derivative with respect to the conformal time, $\tau$. $h_{ij}$ is the traceless transverse perturbations of the metric,
\be
h_{i}^{i}=0\,, \qquad  \partial^i h_{ij}=0\,.
\ee
which means that the tensor perturbations have two helicities. For a near-de-Sitter background,
 \begin{equation}\label{background}
 a(\tau)\simeq -\frac{1}{H\tau},
 \end{equation}
where $H$ is the Hubble constant. The most general solutions to \eqref{u-p-eq}  is of the form of Bessel functions:
\beqa\label{u-h-sol-ds}
u_k(\tau) &\simeq& \frac{\sqrt{-\pi\tau}}{2}\left[\alpha_k^{{}^S} H_{{}_{3/2}}^{{}^{(1)}}(-k\tau)+\beta_k^{{}^S} H_{{}_{3/2}}^{{}^{(2)}}(-k\tau)\right]\,.\\
p_k(\tau) &\simeq& \frac{\sqrt{-\pi \tau}}{2}\left[\alpha_k^{{}^T} H_{{}_{3/2}}^{{}^{(1)}}(-k\tau)+\beta_k^{{}^T} H_{{}_{3/2}}^{{}^{(2)}}(-k\tau)\right]\,.
\eeqa
$H_{3/2}^{(1)}$ and $H_{3/2}^{(2)}$ are the Hankel functions of the first and second kind, respectively, and they behave like the positive and negative frequency modes. The Bogolyubov coefficients
satisfy the Wrosnkian constraint expressed as,
\beqa\label{Wronskian}
|\AS|^2-|\BS|^2=1\,,\\
|\AT|^2-|\BT|^2=1\,,
\eeqa{equation}
 The BD vacuum \cite{Bunch:1978yq} corresponds to $\A=1$ and $\B=0$. However in this work, we assumed that both kinds of perturbations
  can start off from a generic non-BD excited initial state with nonzero $\B$. The scalar and tensor power spectra are defined as
\beqa
{\mathcal P}_{{}_S}=\frac{k^{3}}{2\pi ^{2}}\left| \frac{u_{k}}{z}\right|^2_{{k/{\cal H}\rightarrow 0}}\,.\\
{\mathcal P}_{{}_T}=\frac{k^{3}}{2\pi ^{2}}\left| \frac{p_{k}}{a}\right|^2_{{k/{\cal H}\rightarrow 0}}.
\label{scrpower}
\eeqa
One should be careful that the energy density and pressure carried by the excited initial states do not prohibit inflation.
The energy and pressure density of the fluctuations in excited initial condition  is
\beqa\label{delta-rho-free}
\delta\rho_{{}_{\text{non-BD}}}&\sim& \frac{1}{a(\tau)^4}\int_H^{\infty} \frac{d^3 k}{(2\pi)^3}\ \frac12(|\AS|^2+|\BS|^2-1) k\,,\nonumber\\
\\
\delta p_{{}_{\text{non-BD}}} &\sim& \delta \rho_{{}_{\text{non-BD}}}\,.\nonumber
\eeqa
Noting that $\delta\rho_{{}_{\text{non-BD}}}'\sim \delta p_{{}_\text{non-BD}}'\sim {\cal H} \delta\rho_{{}_\text{non-BD}}$ in the leading slow-roll approximation,
the backreaction will not derail slow-roll inflationary background if
\be\label{background-backreaction}
\delta\rho_{{}_\text{non-BD}}\ll \epsilon\rho_{{}_0}\,,\quad \delta p_{{}_\text{non-BD}}'\ll \eta\epsilon\rho_{{}_0}\,,
\ee
which may be written in terms as
\be
\int_H^{\infty} \frac{d^3 k}{(2\pi)^3} k |\beta^{{}^{S ({\rm or} T)}}|^2\ll \epsilon\eta H^2\mpl^2\,.
\ee
For the mechanism of excitation that we assumed in eq. \eqref{betas-Gaussian}, it turns out that
\begin{eqnarray}\label{beta-scalar-tensor-backreaction}
   \beta_{{}_0}^{{}^{ S({\rm or} T)}} \lesssim \sqrt{\epsilon\eta}\frac{H \mpl}{M_{{}_{ S({\rm or} T)}}^2}.
\end{eqnarray}
As we see, $ \beta_{{}_0}^{{}^{ S ({\rm or} T)}} $ is  inversely proportional to the scale of new physics $M_{{}_{ S({\rm or} T)}} $.

\bibliographystyle{apsrev}

\end{document}